\documentclass{webofc}
\usepackage[varg]{txfonts}   
\usepackage{paralist}
\begin{document}
\title{Usage of GPUs in ALICE Online and Offline processing during LHC Run~3}

\author{\firstname{David} \lastname{Rohr}\inst{1}\fnsep\thanks{\email{drohr@cern.ch}} for the ALICE Collaboration
}

\institute{European Organization for Nuclear Research (CERN), Geneva, Switzerland }

\abstract{%
  ALICE will significantly increase its Pb--Pb data taking rate from the 1\,kHz of triggered readout in Run~2 to 50 kHz of continuous readout for LHC Run~3.
  Updated tracking detectors are installed for Run~3 and a new two-phase computing strategy is employed.
  In the first synchronous phase during the data taking,  the raw data is compressed for storage to an on-site disk buffer and the required data for the detector calibration is collected.
  In the second asynchronous phase the compressed raw data is reprocessed using the final calibration to produce the final reconstruction output.
  Traditional CPUs are unable to cope with the huge data rate and processing demands of the synchronous phase, therefore ALICE employs GPUs to speed up the processing.
  Since the online computing farm performs a part of the asynchronous processing when there is no beam in the LHC, ALICE plans to use the GPUs also for this second phase.
  This paper gives an overview of the GPU processing in the synchronous phase, the full system test to validate the reference GPU architecture, and the prospects for the GPU usage in the asynchronous phase.
}
\maketitle
\section{Introduction}
\label{intro}

ALICE (A Large Heave Ion Collider Experiment~\cite{bib:alice}) is undergoing a major upgrade during the LHC long shutdown 2 in preparation for increasing its data taking rate during LHC Run~3~\cite{bib:aliceupgrade}.
This includes updates to the major tracking detectors ITS (Inner Tracking System~\cite{bib:itsrun3tdr}) and TPC (Time Projection Chamber~\cite{bib:tpcrun3tdr}) as well as an upgrade to the computing scheme within ALICE O$^2$ (Online Offline~\cite{bib:o2tdr}), comprising the three projects EPN (Event Processing Nodes), FLP (First Level Processors), and PDP (Physics and Data Processing).
The readout will no longer be trigger-based but rather ALICE will record 50 kHz of minimum bias Pb--Pb collisions in continuous readout.
Instead of the classical approach with fast online processing for triggers and QA (Quality Assurance) followed by the computing-intensive offline event reconstruction, the processing happens in a synchronous and an asynchronous phase.
In Run~3 all recorded raw data are compressed and stored to an on-site disk buffer during the synchronous phase.
In parallel, the synchronous phase extracts all required data for the detector calibration, followed by an intermediate step that produces the calibration objects.
The asynchronous phase performs the full reconstruction of the data.
It is split between the compute nodes of the online computing farm and the GRID.
Both synchronous and asynchronous phases use the same software and the same infrastructure, but with different settings and thresholds.
The rationale behind using the online computing farm for the asynchronous processing is that the LHC is without beam for a significant amount of time, and on top of that the collision system is pp most of the time, which requires less computing resources than Pb--Pb collisions.
Hence, running only the synchronous processing, the online computing farm would idle or operate at low load for most of the time.

\section{ALICE Data Processing in Run~3}

\begin{figure}[htb]
\centering
\includegraphics[width=0.6\textwidth,clip]{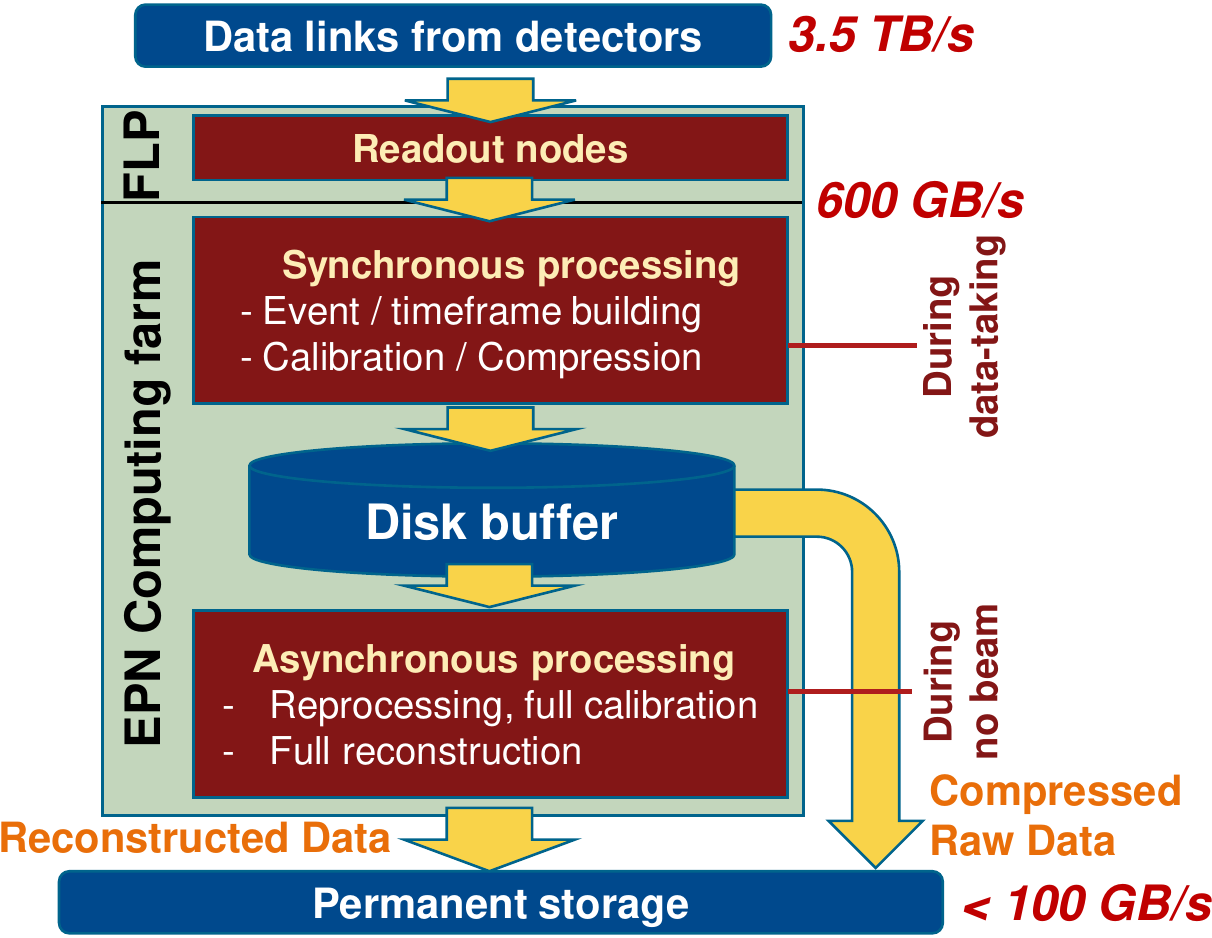}
\caption{Illustration of the ALICE Run~3 processing scheme with two processing phases, synchronous and asynchronous, and the storage to the disk buffer in between.}
\label{fig:o2}
\end{figure}

One general and significant change in the ALICE data processing in Run~3 is the switch from the triggered readout to continuous readout.
In the drift detectors like the TPC, there are no single isolated events, but an overlap of multiple collisions, separated in time, which translates almost linearly to the beamline coordinate.
But since the time interval between collisions is shorter than the TPC drift time, the tracks and the hits originating from the individual collisions cannot be assigned to the vertices a priori.
Instead, a larger bunch of continuous data spanning multiple drift times is reconstructed together, and the assignment of tracks to collisions happens only after the track reconstruction.
The basic processing unit for ALICE in Run~3 is thus not the event, but the Time Frame (TF), which contains by default 128 LHC orbits (around 10\,ms) of continuous data.
The processing scheme is illustrated in Fig.~\ref{fig:o2}.

\subsection{Synchronous Processing}

The synchronous processing has two main objectives.
First, the compression of the raw data to Compressed Time Frames (CTF), which are stored to the disk buffer.
And second, the extraction of all data required for the calibration, such that the calibration tasks may run afterwards without accessing the raw data again.

\looseness=-1
The incoming raw data rate at 50\,kHz Pb--Pb is over 3 TB/s, thus it is impossible to store the data, even temporarily.
Also a transfer over a longer distance would be complicated.
Instead, it must be compressed in real time to a manageable size in the on-site online computing farm.
The farm is split in two parts.
The data enters the farm in the First Level Processors (FLP) via the Common Readout Units (CRU)~\cite{bib:cru}.
The CRUs are FPGA-based PCI Express (PCIe) cards that receive the detector data via optical links, optionally perform local processing in the user logic in the FPGA, and transfer the data to the memory of the host.
Each FLP sees only the data from the links connected to it, i.\,e.~only parts of individual detectors, but it has access to all data arriving from that link.
From the FLPs the data is sent over an Infiniband network to the Event Processing Nodes (EPN).
The EPN is the larger computing farm and provides the majority of the computing capacity.
The event building happens during this network transfer, which is arranged such that each EPN receives full time frames, i.\,e.~all data from all links for an individual time frame.
Thus, the EPN sees the data from all detectors, but only for the duration of one time frame.
Calibration tasks can run on both the FLPs and the EPNs.
By design, tasks operating on link-level, which need access to the continuous data stream of a link, i.\,e. more than one time frame, must run on the FLP.
Other tasks will generally run on the EPN, particularly those that need global information.

Data compression happens in both the FLPs and the EPNs.
The detector with the by far largest data volume is the TPC, contributing with more than 90\% to the total data size.
The CRUs in the FLPs perform zero-suppression of the TPC data before shipping it to the EPNs, reducing the rate from 3.5\,TB/s to 600\,GB/s.
The EPNs further compress the data down to a rate of around 100\,GB/s, which is then stored to the disk buffer as CTFs.
A common lossless ANS~\cite{bib:ans} entropy encoding is employed as the last stage for all detectors, and individual detectors may employ specialized prior compression steps.

\looseness=-1
The TPC compression is the most elaborate one, involving several steps~\cite{bib:ctd2019}, some of which are not lossless.
In particular, a clusterizing algorithms converts the raw ADC values to hits, those are stored in custom data formats with only as many bits as necessary to match the intrinsic TPC resolution.
Hits of tracks not used for physics analysis are removed, while the remaining hits are processed by entropy-reduction steps such as the track model compression~\cite{bib:lhcp2017,bib:ctd2019}.
By design the identification of the hits to be removed requires full tracking in the TPC during the synchronous processing as prerequisite.

\looseness=-1
The most computing-intensive task of the calibration is the TPC space-charge distortion (SCD) calibration~\cite{bib:lhcp2017,bib:ctd2020}, which requires matching and refitting of ITS, TPC, TRD (Transition Radiation Detector), and TOF (Time of Flight) tracks.
This requires track reconstruction for several detectors, but with the increased Run~3 interaction rate, processing in the order of one percent of the events is enough for the calibration.
Consequently, the TPC tracking for all collisions, as required for the data compression, is the dominant workload of the synchronous processing.

\subsection{Asynchronous processing}

In contrast to the synchronous processing, the asynchronous processing includes the reconstruction for all detectors, and for all events instead of only a subset.
Therefore, the processing workload for all other detectors except the TPC is much higher than during the synchronous processing.
For the TPC, in contrast, the clustering and the data compression is not necessary during the asynchronous processing, while the tracking runs on a smaller input data set, as a subset of the hits were removed during the data compression.
Consequently, the TPC processing needs less time in the asynchronous phase than in the synchronous phase.
Overall, the TPC contributes a significant part to the asynchronous processing, but it is not dominant.

\looseness=-1

In contrast to synchronous processing of raw data, which must happen exclusively on the EPNs, the CTFs are available in the GRID.
The asynchronous reconstruction will be split between the EPN farm and the GRID sites.
While the final distribution scheme is still to be decided, currently it is assumed to split it to around one third for the online computing farm, the Tier 0, and the Tier 1 sites each.

\subsection{The Event Processing Nodes (EPN) online computing farm}

The EPN farm is tailored for the fastest possible TPC track reconstruction, which is the bulk of the synchronous processing.
GPUs have been shown to excel at this task in the ALICE High Level Trigger (HLT) during Run~2~\cite{bib:hltpaper}, hence the EPN farm provides the majority of its computing power in the form of GPUs.
The initial version of the O$^2$ tracking was derived from the HLT tracking, and it has been improved in many aspects since then~\cite{bib:ctd2018}.
These improvements include in particular a better track resolution matching the Run~2 offline resolution, the treatment of data from continuous readout with overlapping collisions in the TPC and the absence of absolute, a priori $z$-coordinates, SCD corrections, and efficiency improvements to function with the five-fold increase of the TPC occupancy compared to Run~2.
Since the data cannot be buffered, the EPN computing capacity must be sufficient for the highest data rates expected during Run~3, which is 50\,kHz of Pb--Pb collisions.
Thus, the TPC processing speed together with the maximum TPC input data rate define the number of GPUs in the farm.
The rest of the farm is composed to ensure maximum GPU performance and minimal cost for the server infrastructure.
For instance the number of required servers is a function of the number of GPUs and of how many GPUs fit in one server.
Plugging more GPUs in the same server reduces the number of servers enabling some cost-saving for the overall farm.
The reference configuration for the EPN computing farm are dual-socket servers with 8 GPUs each.

During the synchronous phase, the GPUs will be fully loaded by the TPC reconstruction.
With an EPN farm providing 90 percent of its compute performance via GPUs, it is desirable to maximize the GPU utilization also in the asynchronous phase.
Since the relative contribution of the TPC to the overall workload is much smaller in the asynchronous phase, GPU idle times would be high and processing would be CPU-limited, if only the TPC part would run on the GPU.
Therefore, ALICE attempts to port more computing-intense reconstruction steps onto the GPUs.

Still, the primary objective of the EPN farm, and also of the software effort, is to enable synchronous data taking and processing at 50\,kHz Pb--Pb rate.
If the EPN farm is unable to sustain the full rate, there is no fallback solution, and ALICE would inevitably lose data.
Therefore, the GPU software was developed with two scenarios in mind:
\begin{compactitem}
  \item \textbf{Baseline scenario:} This mandatory scenario contains the GPU processing steps required to ensure synchronous processing at 50\,kHz Pb--Pb rate.
  In addition, processing steps with either already existing or nearly final GPU versions (e.\,g.~for ITS~\cite{bib:itsgpu}), which enables the GPU usage with little development effort, are included.
  \item \textbf{Optimistic scenario:} Here, the goal is to run as many components as possible on the GPU.
  Obviously, processing steps with negligible computing workload should not be ported, and porting individual short processing steps, which would require data transfers forth and back, is avoided.
  Ideally, a chain of many consecutive computing intense workloads should run on the GPU.
  The full track reconstruction for all detectors in the central barrel is considered as a good candidate, but the GPU effort includes other detectors as well.
\end{compactitem}
\begin{figure}[htb]
\centering
\includegraphics[width=\textwidth,clip]{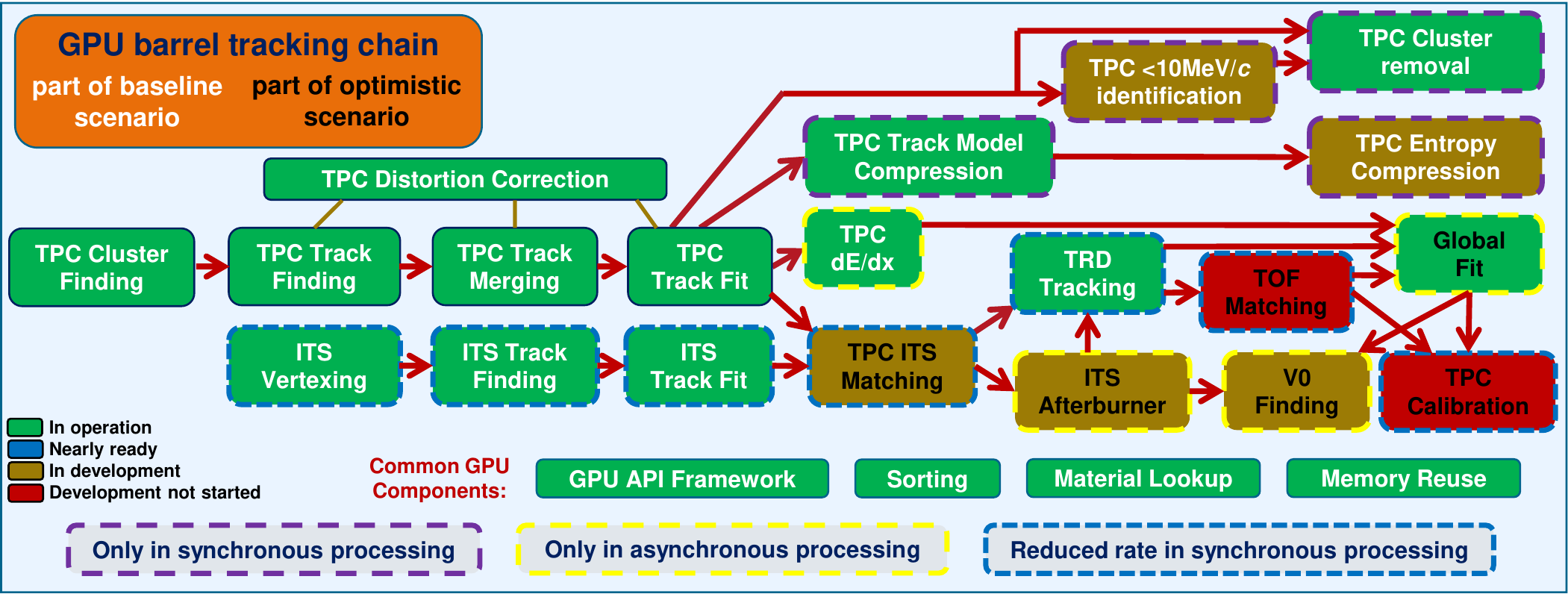}
\caption{Illustration of the major processing steps in the ALICE central-barrel global-tracking chain.
The development status of each step is indicated by the background color.
The text color indicates whether the step is part of the baseline or optimistic scenario.
The box borders highlight steps which run either only in the synchronous, or only in the asynchronous phase, or which run in both phases but process only a fraction of the data during the synchronous phase.}
\label{fig:schedule}
\end{figure}
Figure~\ref{fig:schedule} shows the relevant processing steps of the central-barrel global-tracking chain, and highlights which components belong to the synchronous  or the asynchronous phase or to both, whether they run only for a reduced subset of the collisions, whether they belong to the baseline or the optimistic scenario, and the current state of the GPU development.
All GPU code is developed in a generic way that can target different GPU architectures and can also run on the CPU~\cite{bib:generic}.
Hence, all the components shown in Fig.~\ref{fig:schedule} can run on the CPU as well.
It is important to note that there are working CPU-only versions of the components not yet ported to GPU, i.\,e.~the tracking chain as shown is fully operational on the CPU, or in a hybrid mode running the GPU-enabled parts on the GPU and the rest on the CPU.

\section{Performance evaluation and testing}

This section gives an overview of the benchmarking performed on different hardware platforms, the constraints for the final hardware, and the final full system qualification test.
Extensive benchmarks of several GPU types of the AMD Vega family and the NVIDIA Pascal, Turing, and Ampere families were performed.
The GPU employed eventually in the EPN is the AMD MI50.

\subsection{Constraints}

\subsubsection{CPU cores}
During GPU processing, some CPU capacity is needed for feeding the data to the GPU, controlling the GPU, and also for some minor CPU processing steps.
On top of that, there are the CPU resources for the actual CPU reconstruction steps, which are not considered here.
The ALICE GPU code is written such that a single CPU core performs all tasks for one GPU, also when using multiple command queues.
Even though this core is sometimes not at 100\% load, it is counterproductive to place additional work on this core, since it reduces the CPU responsiveness to GPU events, thus decreasing the GPU performance.
Consequently, each GPU needs one full CPU core for the user code, while additional load may come from the GPU driver and runtime, depending on the platform.
With NVIDIA CUDA, this overhead for the ALICE applications is absolutely negligible, while with AMD HIP an additional CPU load of around 1.5 CPU cores per GPU is observed.
Thus, for operating eight AMD GPUs in a server, at least 20 CPU cores are required to perform the GPU tasks, while eight cores are required in the NVIDIA case.

\subsubsection{PCI Express Bandwidth}
The ALICE TPC code (which transfers the most data) is written with a pipeline such that the transfer of input and output data overlaps with the processing~\cite{bib:hltpaper}.
As long as the PCIe transfer is fast enough, the GPU will never idle.
The average data rates sent over PCIe are rather low, but it must be considered that there is a peak load during the TPC clusterization phase, when all the TPC raw data must be shipped to GPU.
In order to avoid stalling the GPU, PCIe must keep step with this particular transfer.
The required speed depends on the GPU performance.
The impact of different PCIe generations on the performance was measured to be negligible, as long as the processing pipeline can fully hide the asynchronous PCIe transfer behind the processing.
For the MI50, even PCIe Gen 2 would be sufficient, showing an overall performance decrease of around 1\% vs.~Gen 3 with no difference between Gen 3 and Gen 4.
Some other GPUs require Gen 3 for the full performance.

\subsubsection{Memory size}
\begin{figure}[htb]
 \begin{minipage}[t]{0.485\textwidth}
 \centerline{\includegraphics[width=0.955\textwidth]{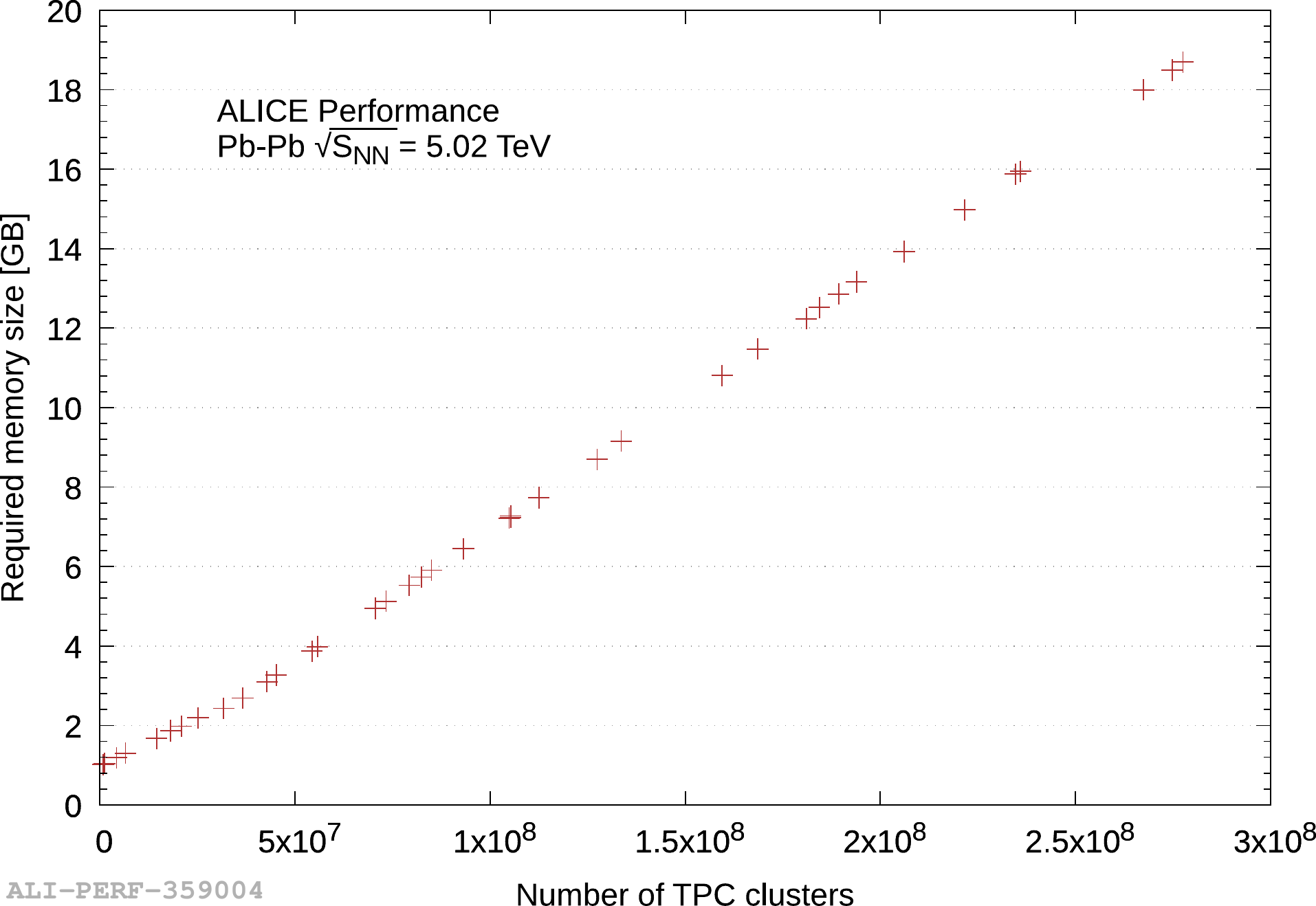}}
 \caption{Maximum GPU memory consumption during synchronous reconstruction vs.~input data size.}
 \label{fig:memory}
 \end{minipage}
\hfill
 \begin{minipage}[t]{0.485\textwidth}
 \centerline{\includegraphics[width=0.955\textwidth]{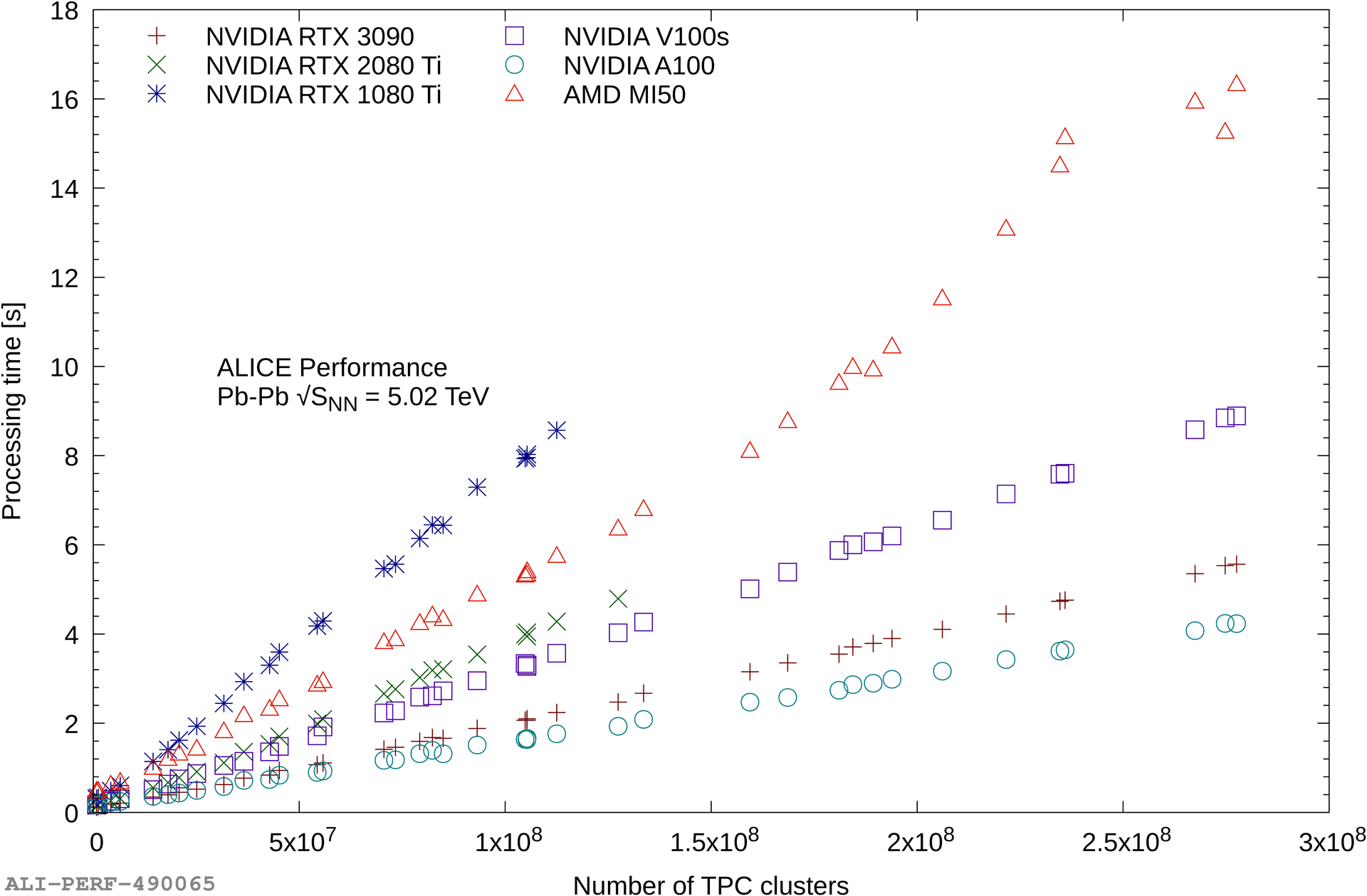}}
 \caption{Processing time of the synchronous reconstruction on different GPUs vs.~input data size. Note that some GPUs are memory-limited.}
 \label{fig:perf}
 \end{minipage}
\end{figure}

\looseness=-1
Two memory constraints must be considered.
Since ALICE always processes a full time frame at once, the full time frame must fit in GPU memory.
Much effort is put into reusing GPU memory for consecutive processing steps~\cite{bib:chep2019}, but at all times the raw data and the scratch data of one processing step must fit.
Figure~\ref{fig:memory} shows the memory requirements.
The rightmost data points are for time frames with the default length of 128 LHC orbits and with the occupancy of 50 kHz Pb--Pb.
It is not planned to run with higher occupancies, but we have verified that the TPC reconstruction performance behaves identically up to 256 LHC orbits.
Note that on top of the memory used by the application, the GPU runtime requires additional memory, e.\,g.,~for the stack.
It must also be considered that due to fluctuations in centrality and luminosity, the number of TPC hits and thus the required memory size varies to a small extent, demanding a certain margin.
Taking this into account, a 24 GB GPU is sufficient in the vast majority of cases and only less than 0.1\% of the time frames would not fit.
Such  time frames could easily be processed on the CPUs.
The actual EPN farm will be equipped with 32 GB GPUs, which do not have any memory limitation.

\looseness=-1
The potential reduction of the TF length was also studied..
At a length of 70 orbits, 16 GB GPUs would be sufficient.
As is shown in the next section, the time frame length has no general impact on the performance, as long as a certain minimum length of around 5 orbits is exceeded.
The processing time, for the most part, goes linearly with the input data size.
However, there can be certain non-linear effects for certain GPU models at certain sizes, such as cache effects.
This must be verified for a particular model and such sizes should be avoided.
Since time frames are processed individually, and at the end of each time frame there are collisions with the TPC drift time reaching into the next time frame, there are a small number of collisions that cannot be reconstructed~\cite{bib:chep2019}.
There are technical solutions to recover these collisions by storing a small amount of additional data, but in order to keep things simple and to minimize the data loss, it was decided to stick to the 128 orbit time frame length.

Besides the GPU memory, there are also requirements on the host memory.
With eight GPUs, the host must hold at least 8 times the input data and buffers for the output data in memory.
In addition, to feed the GPU pipeline and to enable further pipelined processing of the output on the CPU, an input queue containing four input time frames waiting in memory and an output queue of eight time frames for postprocessing by CPU tasks are foreseen.
With around 5\,GB input data size and around 10\,GB output data size, this sums up to 220 GB of memory.
Adding memory for the CPU processing steps, network buffers, the operating system, etc., it is clear that 256\,GB of system memory will be insufficient, while 512\,GB should be enough (compare the measurements in section~\ref{sec:fullsystem}).
It was not considered to use a non-power-of-two memory size, to equally load all 16 memory channels with identical modules.

\subsection{Standalone benchmark}

\label{sec:standalone}

\begin{figure}[htb]
 \begin{minipage}[t]{0.485\textwidth}
 \centerline{\includegraphics[width=0.955\textwidth]{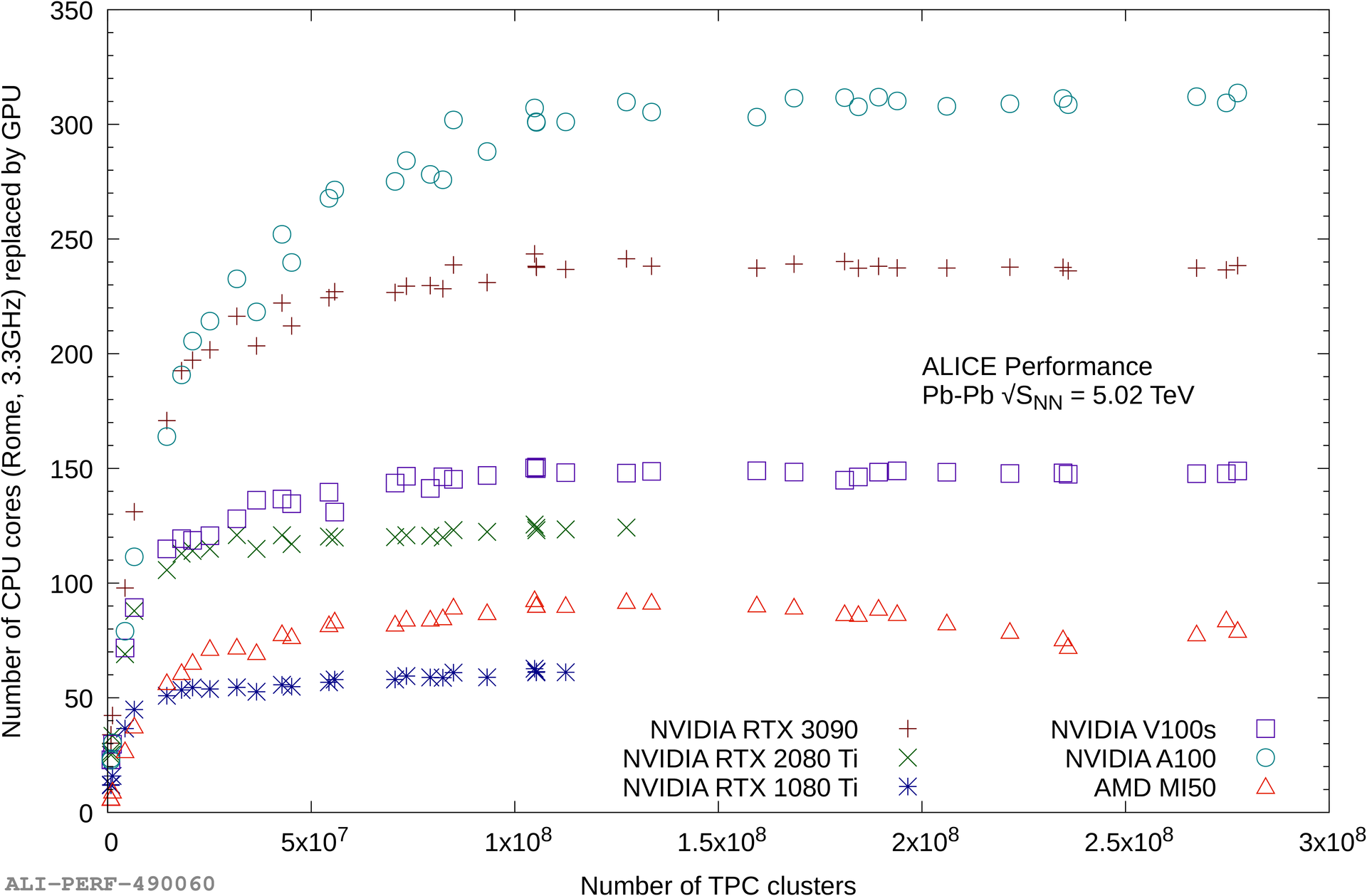}}
 \caption{Speedup of different GPU models over a single AMD Rome core running at 3.3 GHz vs.~input data size (refer to Fig.~\ref{fig:cpuscaling} for the CPU multi-core performance).
 It is corrected for the CPU load caused by the GPU processing, i.\,e.~the average number of CPU cores used during the GPU processing is subtracted from the speedup.}
 \label{fig:speedup}
 \end{minipage}
 \hfill
 \begin{minipage}[t]{0.485\textwidth}
 \centerline{\includegraphics[width=0.955\textwidth]{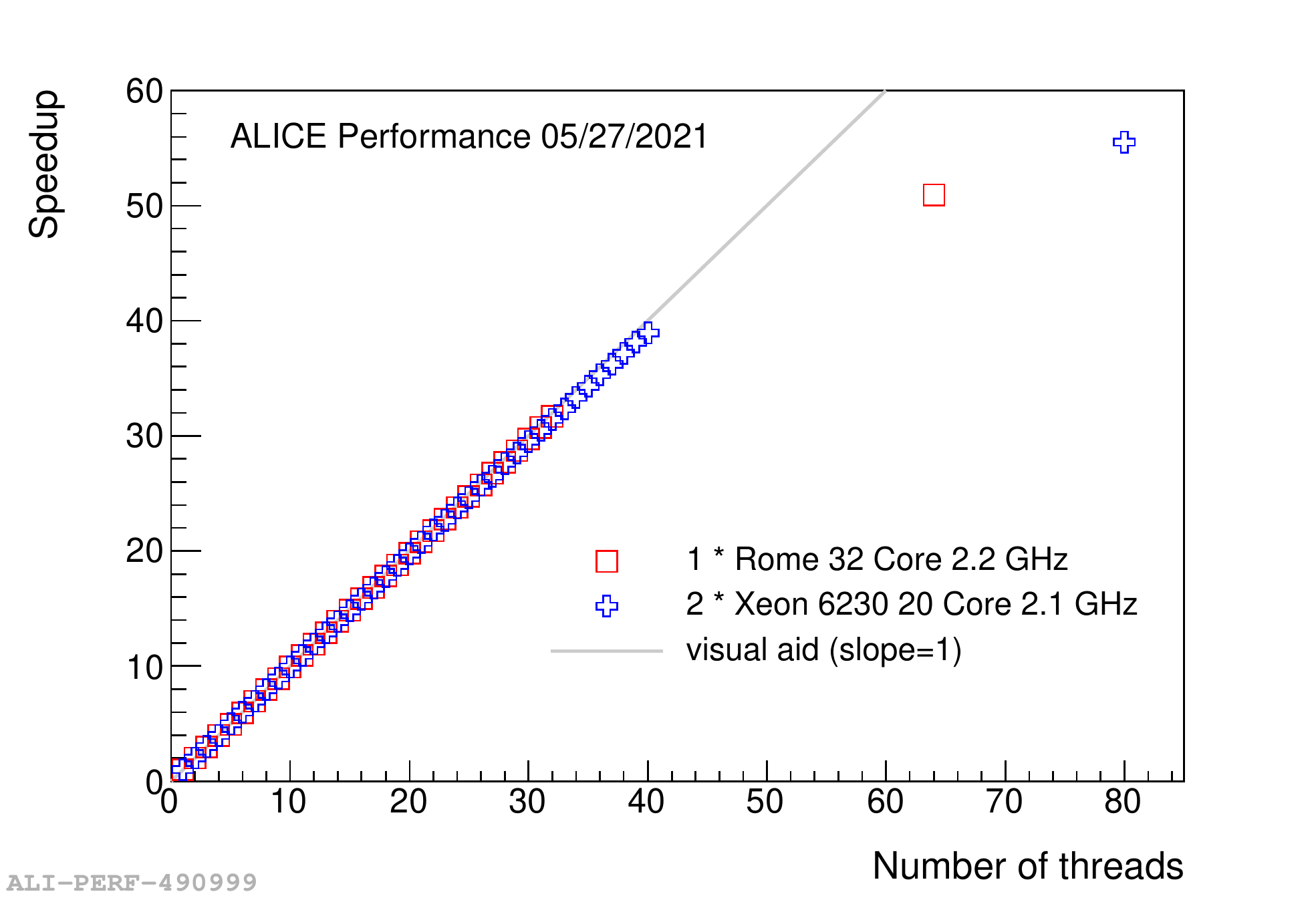}}
 \caption{Speedup of multi-core processing on AMD and Intel CPUs.
 Thread placement is even over CPU sockets and AMD CPU chiplets.
 HyperThreaded / SMT cores are not used except for the rightmost measurement points, which uses all cores.
 CPU frequencies were pinned to the indicated values.}
 \label{fig:cpuscaling}
 \end{minipage}
\end{figure}

\looseness=-1
Performance is first evaluated by a standalone benchmark, which loads time frames in the raw input format into memory, and processes them on the GPU in an endless loop as fast as possible, without any additional framework code.
The standalone test uses only a single GPU, but it was tested to run eight GPUs in the same server independently in parallel, which revealed an efficiency loss of less than 1\%.
Despite possible bottlenecks from PCIe transfers or CPU processing, this should achieve 100\% GPU load, while in reality around 99\% GPU load is achieved.
Figure~\ref{fig:perf} shows the processing time of different GPU models against the input data size measured in number of TPC hits.
The speedup compared to a single CPU core is shown in Fig.~\ref{fig:speedup}.
It should be noted that these numbers are for the synchronous reconstruction only, and are a bit biased towards the GPU since the employed clusterizing algorithm performs much better on GPUs than on CPUs.
For comparison, the speedup of the MI50 GPU in the asynchronous reconstruction is in the order of 50 CPU cores.
The comparison was made to a single CPU core due to the vast majority of CPU models and core counts.
For comparing with multi-core processing on the CPU, one can practically scale linearly with the number of cores.
Figure~\ref{fig:cpuscaling} shows that on both AMD and Intel platforms the scaling to multiple CPU cores is almost linear with an efficiency greater than~98\% for 32 and 40 cores respectively.
It also shows a significant improvement by enabling HyperThreading / SMT, which is thus active on the EPN farm.

\begin{figure}[htb]
 \begin{minipage}[t]{0.485\textwidth}
 \centerline{\includegraphics[width=0.955\textwidth]{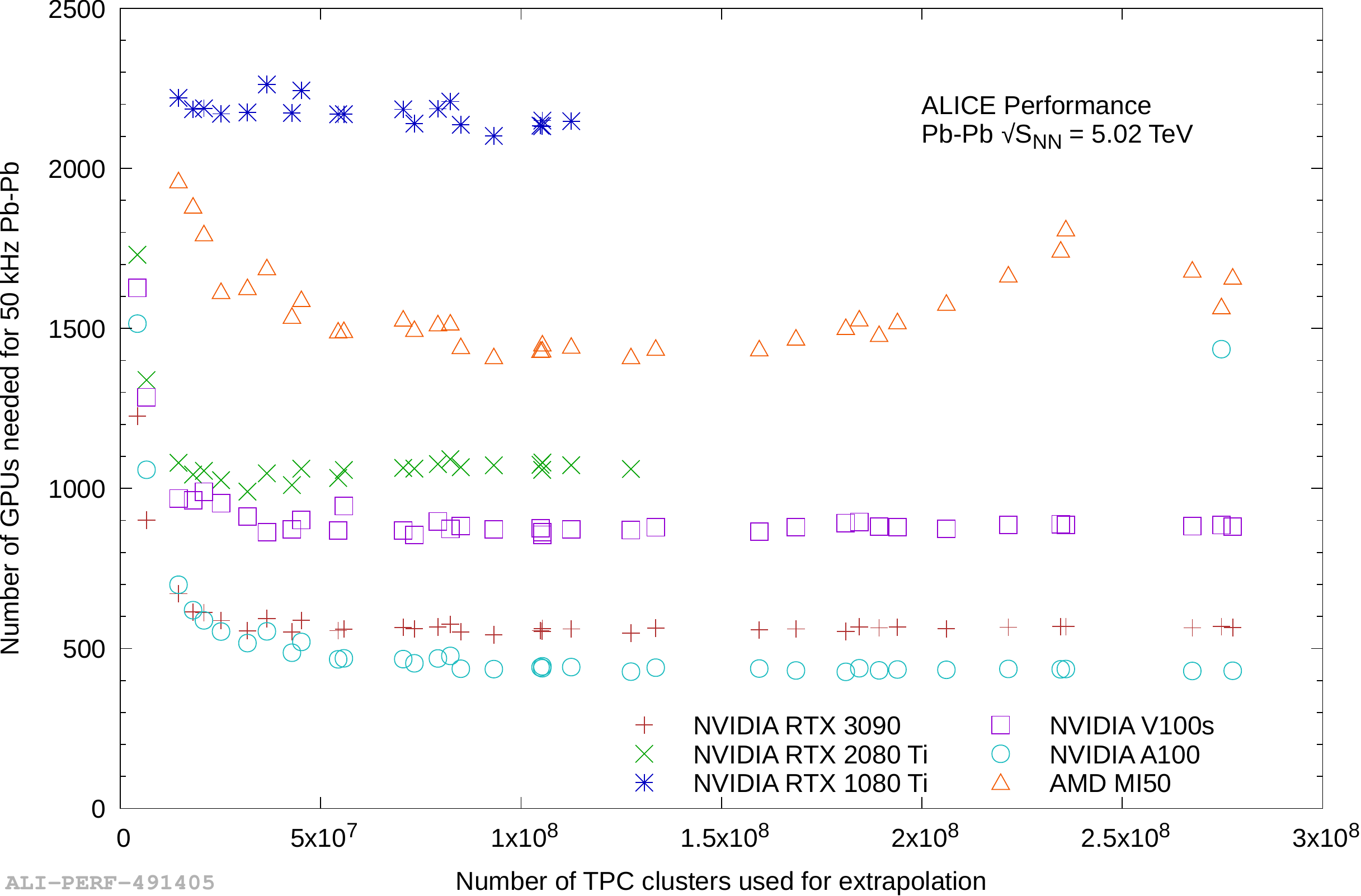}}
 \caption{Number of GPUs needed (computed as processing time per time frame / time frame length) for the synchronous reconstruction vs.~input data size.
 Derived from the standalone benchmark for 50\,kHz Pb--Pb rate for GPUs operating at steady 100\% load.
 The measured processing time obtained for a time frame (values of Fig.~\ref{fig:perf}) is scaled linearly by the number of the TPC hits to the average number of hits in a time frame to correct for centrality and luminosity fluctuations.}
 \label{fig:gpus}
 \end{minipage}
 \hfill
 \begin{minipage}[t]{0.485\textwidth}
 \centerline{\includegraphics[width=0.955\textwidth]{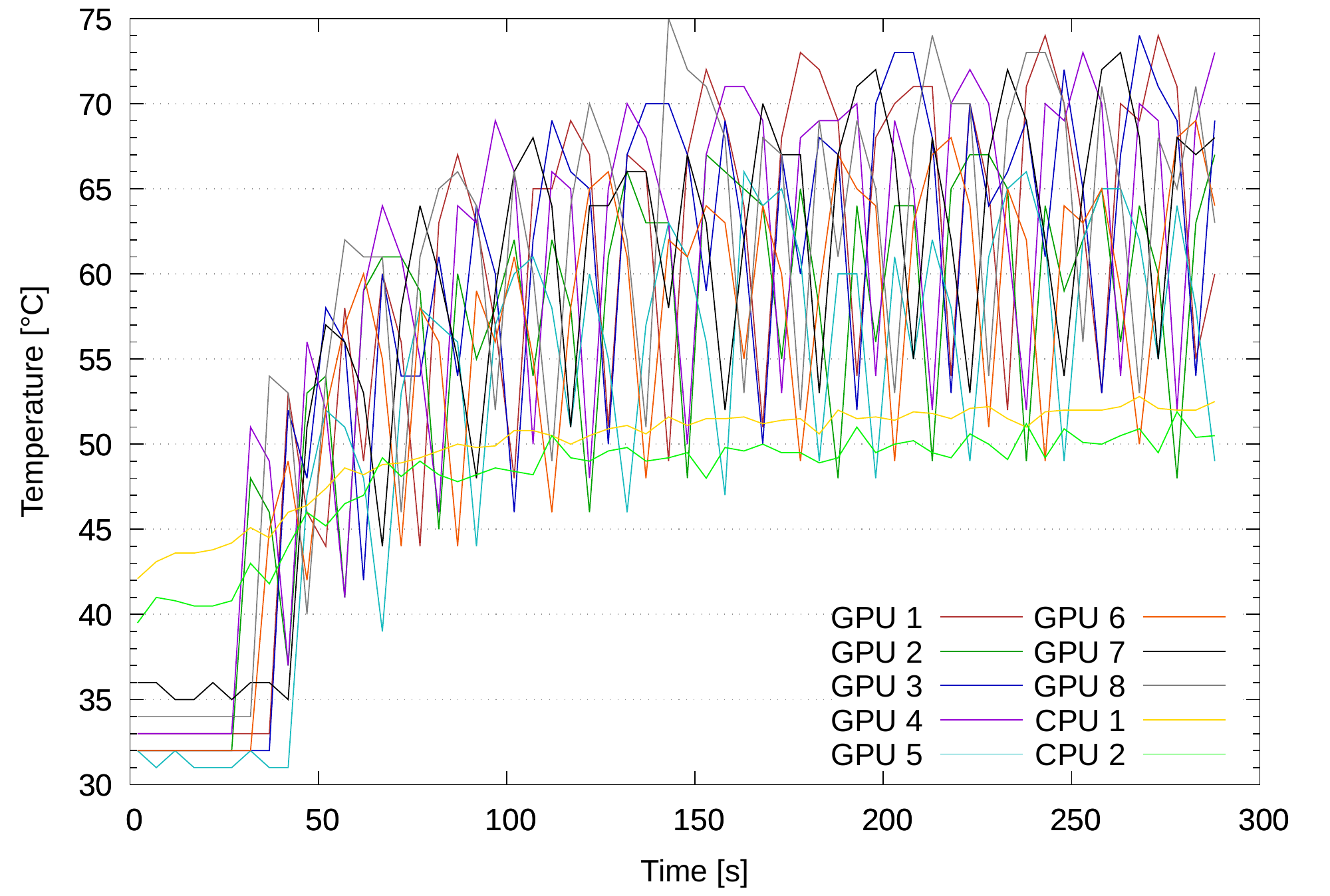}}
 \caption{GPU temperatures measured during the full system test at 21°C environment temperature.
 The strong fluctuations are due to the GPUs cooling down during the short idle periods of~1 to~3\,seconds in the full system test after around~10 seconds of full load for processing each time frame.}
 \label{fig:temp}
 \end{minipage}
\end{figure}

From these numbers one can compute how many GPUs are required to cope with the 50\,kHz Pb--Pb interaction rate, which is shown in Fig.~\ref{fig:gpus} for different GPU models.
Please note that this estimate assumes continuous 100\% GPU load, which is unrealistic in a real time system.
There will be inevitable latencies, and assuming 100\% load, the farm could never catch up any latencies that will appear.
In addition, a margin is needed for two more reasons:
\begin{compactitem}
 \item There will be further improvements to the physics performance of the tracking, and unfortunately better performance does not imply faster code in this case.
 To the contrary, if the track finding and cluster attachment efficiencies increase, the GPU has to fit more tracks with more clusters, and the processing time will increase almost linearly.
 \item The upgraded TPC is a new detector that was never operated under 50\,kHz Pb--Pb conditions yet and thus not all parameters are exactly known.
\end{compactitem}

\looseness=-1
Contrarily, there is also further possibility to speed up the code.
Overall, a margin of 20\% over the estimate of Fig.~\ref{fig:gpus} is considered sufficient.
In the case of the AMD MI50 model of the actual EPN farm, this means around 2000 GPUs and thus 250 servers are required.

\subsection{Full system test}

\label{sec:fullsystem}

\looseness=-1
In addition to the standalone benchmarks, a full system test of an EPN server prototype was conducted in preparation of the EPN Production Readiness Review (PRR).
The test was conducted on a prototype server with two 64 core Rome CPUs, out of which two times 32 cores were disabled in the BIOS to match the reference configuration.
The server is equipped with 512 GB of RAM, one Infiniband HCA, and eight AMD MI50 GPUs.
Due to GPU availability at that time, the 16 GB model was used, which limited the time frame length to 70 orbits for the test.
For validation, the same test was also successfully performed on a server with two NVIDIA 2080 Ti GPUs.
The full system test is limited to a single server.
However, since all servers are identical and are processing individual time frames independently, there is no difference compared to testing multiple servers, except that it takes more time to obtain an acceptable statistical precision.
Since no setup with enough FLP servers providing the input was available, time frames were replayed from within the EPN memory without event building.
The CPU overhead for the network transfer and event building was measured independently to be six CPU cores or less.
This number is added to the CPU load measurements of the full system test.

The biggest difference to the standalone benchmark is that the GPUs are not processing time frames as fast as possible, but time frames are published at a fixed rate and are then processed by the full synchronous tracking chain including the CPU components (see Fig.~\ref{fig:schedule}).
It should be noted that some CPU components for some detectors were not ready at the time of the test.
Conservatively estimated, the included components account for at least 80\% of the total CPU processing workload, thus 25\% is added on top of the CPU processing workload measured in the full system test.
The distribution scheme at a fixed rate means that the GPUs will not be loaded at 100\%, but close to it if the rate is chosen aggressively.
The results shown below are for a rate that corresponds to $\frac{1}{250}$ of the 50\,kHz Pb--Pb rate, which will reach each single server out of the 250 server farm.
The tests were repeated with $\frac{1}{230}$ of the rate to demonstrate a sufficient margin, which worked without any problems while CPU load and buffer usage increased accordingly.

Figure~\ref{fig:temp} shows the GPU temperatures measured throughout the tests, which never exceeded 75$^\circ$\,C, leaving 15$^\circ$ headroom to the 90$^\circ$\,C temperature at which the GPUs begin to throttle.

\begin{figure}[htb]
 \begin{minipage}[t]{0.485\textwidth}
 \centerline{\includegraphics[width=0.955\textwidth]{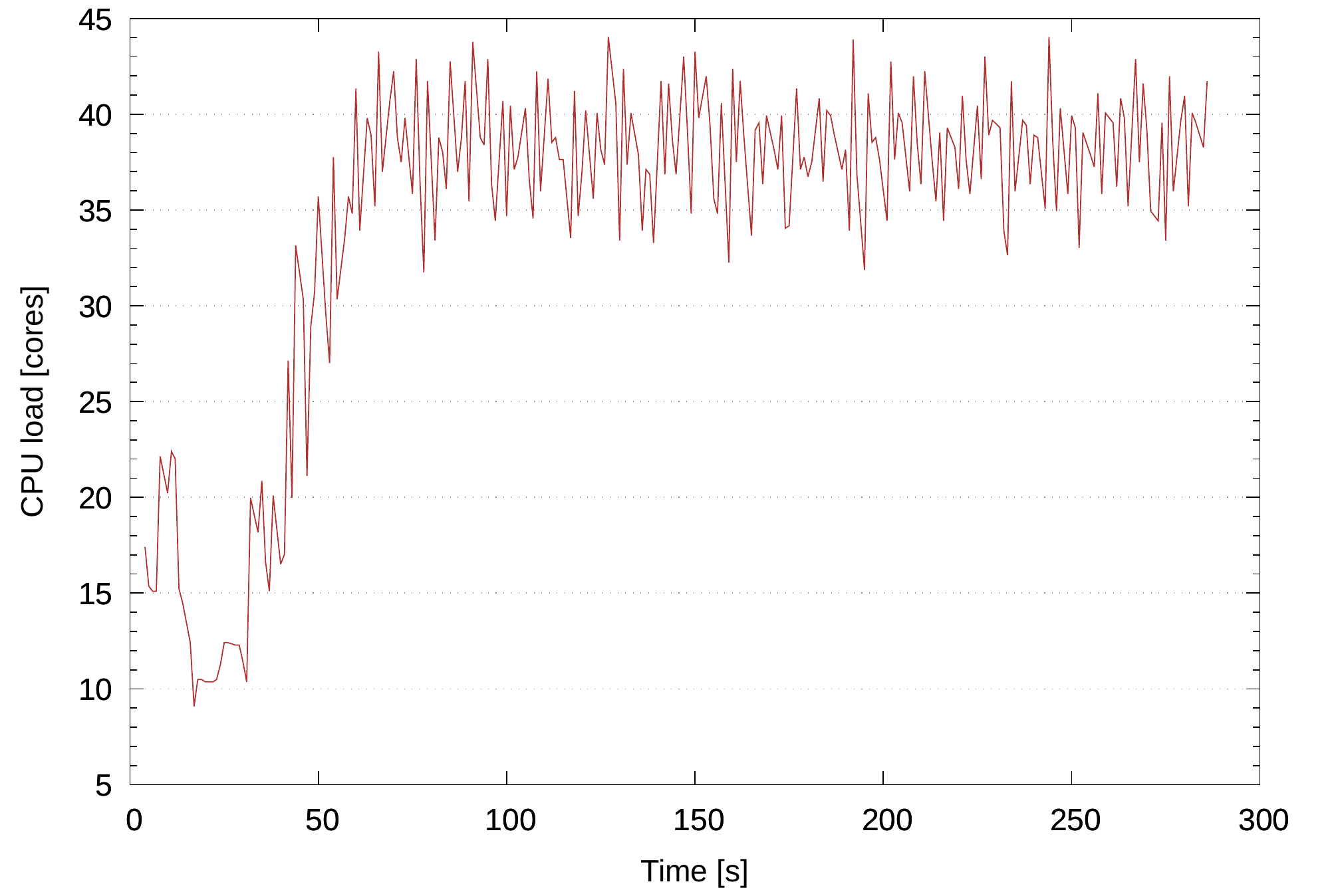}}
 \caption{Number of CPU cores used during the first 5 minutes of a full system test with eight GPUs at $\frac{1}{250}$ of the 50\,kHz Pb--Pb data rate.
 The load is smaller during the startup and then oscillates around 38 cores due to small fluctuations of the load of the CPU parts of the reconstruction.}
 \label{fig:cpuuse}
 \end{minipage}
 \hfill
 \begin{minipage}[t]{0.485\textwidth}
 \centerline{\includegraphics[width=0.955\textwidth]{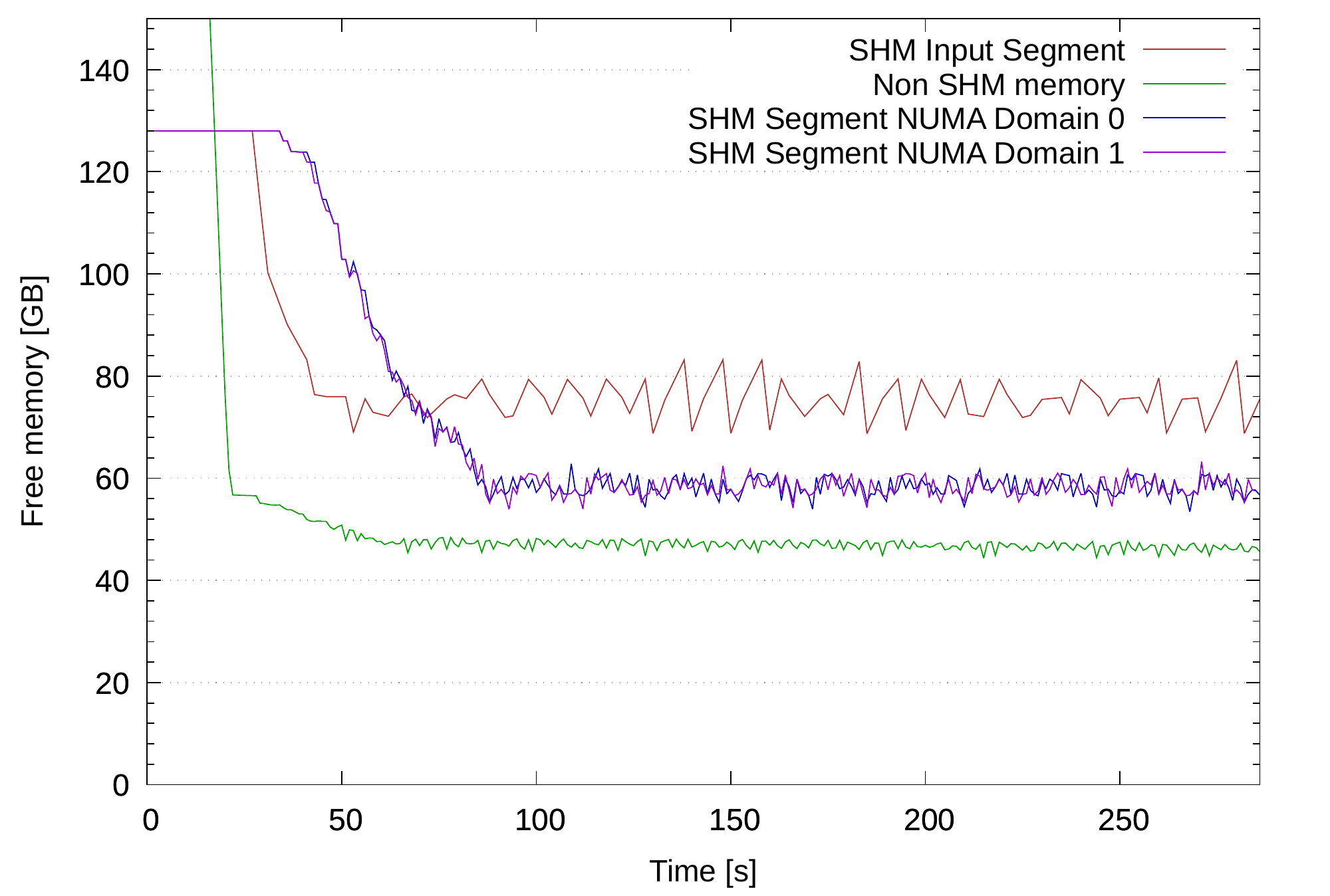}}
 \caption{Memory usage during the first 5 minutes of a full system test with 8 GPUs at 50\,kHz Pb--Pb rate.
 The free memory in the four distinct memory regions is shown.
 After around 90 seconds into the test, the memory utilization is fully stable.}
 \label{fig:memuse}
 \end{minipage}
\end{figure}

Figure~\ref{fig:cpuuse} shows the CPU usage during the full system test measured as number of cores used.
The maximum CPU load was 44 cores.
Out of these, 25 cores were used for the CPU processing steps, which must be inflated by 25\%.
Adding six cores for network transfers and event building on top, the full system test demonstrates that 56 cores are sufficient for the synchronous processing.
The EPN farm employs two 32 core processors, having a margin of eight cores, and all SMT cores on top.

Figure~\ref{fig:memuse} shows the memory usage during the full system test.
The memory is split in four regions.
There are two Shared Memory (SHM) regions, one per NUMA (Non Uniform Memory Architecture) domain, which are used for the data exchange between the various processing components.
Basically the EPN operates as two virtual EPNs with four GPUs each, one per NUMA domain, however with a common input by the one Infiniband network adapter.
A third shared memory region in interleaved memory is used for the input.
The remaining non-SHM memory of the host is available as scratch memory for the processes.
Currently, one disadvantage of this approach is that free memory cannot be moved at runtime between the regions.
Thus each region must maintain a margin of free memory.
The total free memory of the server is the sum over the four regions, but if one region runs out of memory, the processing will stall and create back pressure, until some memory is freed.

The natural way to verify that the server can cope with a certain rate is to observe the memory buffers.
If the fixed rate exceeds the processing capacity, the published input data will start fill up the input buffers and the remaining buffer size will go to zero causing back pressure.
If the memory usage is constant over time, the processing speed is sufficient.
This is the case for an EPN server processing $\frac{1}{230}$ of the 50\,kHz Pb--Pb rate.

\section{Conclusions}

The ALICE data processing scheme for Run~3 was described and the results of the full system test conducted for the synchronous event reconstruction presented.
The baseline solution for using GPUs in the computing-intense workloads of the synchronous processing is fully implemented and established.
The 250 servers of the reference configuration, each equipped with eight AMD MI50 GPUs, two 32-core Rome CPUs, and 512 GB of memory, are capable of handling the Pb--Pb interaction rate of 50 kHz with roughly 20\% margin.
One GPU of the employed model replaces around 50 CPU cores.
On top of that, ALICE is aiming for a more optimistic scenario to port the full central-barrel tracking chain to the GPU.
This effort is ongoing, but already several components beyond the synchronous reconstruction baseline are available on the GPU.
In parallel, other subdetectors not taking part in the central-barrel tracking are investigating GPU adoption independently.
All GPU code is written in a generic way, such that GPU platforms can be switched easily, and the same algorithms run also on the CPU.

\end{document}